# Parametric instability of non-Hermitian systems near the exceptional point

A.A. Zyablovsky,[1,2] E.S. Andrianov,[1,2] A.A. Pukhov[1,2,3]

[1] All-Russia Research Institute of Automatics, 22 Suschevskaya, Moscow, Russia
[2] Moscow Institute of Physics and Technology, 9 Institutsky per., Dolgoprudny, Russia
[3] Institute Theoretical and Applied Electromagnetics, 13 Izhorskay per., Moscow, Russia

*Abstract*

In contrast to Hermitian systems, eigenstates of non-Hermitian ones are in general non-orthogonal. This feature is most pronounced at exceptional points where several eigenstates are linearly dependent. In this work we show that near this point a new effect takes place. It exhibits in energy increases in the system when its parameters change periodically. This effect resembles parametric resonance in a Hermitian system but there is a fundamental difference. It comes from the unique properties of the exceptional point that leads to parametric instability that occurs almost at any change in a parameter, while in the case of Hermitian systems it is necessary to fulfill resonance conditions. We illustrate this phenomenon by the case of two coupling waveguides with gain and loss. This phenomenon opens a wide range of applications in optics, plasmonics, and optoelectronics, where the loss is an inevitable problem and plays a crucial role.

*Introduction*

Recent developments in nanoscience have led to dramatic decreases of system size. The inherent properties of such small-size systems are the impossibility of separation of the environment from the system under consideration. Interaction with the environment leads to the dissipation of energy in the system. For this reason, the investigation of open and, in particular, non-Hermitian systems has been the main topic of physics in the last decade [1-21].

Besides the problem of loss compensation, non-Hermitian systems attract attention due to their specific properties which do not occur in Hermitian ones. In general cases, the eigenstates $|\varphi\rangle$ of non-Hermitian systems are not orthogonal [2, 3, 22], unlike those of Hermitian ones. The maximal degree of non-orthogonality is reached at the exceptional point (EP), where some of the eigenvectors become linearly dependent [3].

Non-orthogonality of eigenstates leads to non-reciprocal propagation and energy oscillations when the total energy in the system oscillates in the propagation in waveguides [2, 22]. Also, due to the non-orthogonality of eigenstates, non-Hermitian systems may exhibit non-exponential transient behavior [23] when all the eigenstates are decaying. In this case the field amplitude increases compared with the initial value in the first stage and decreases exponentially in the second stage [23]. Among other non-Hermitian systems the PT-symmetric ones are under active consideration [4-6]. An unique property of these systems is their spectrum may be both real and complex [1-6]. In PT-symmetric systems, the EP and the point of phase transition from real spectrum to complex one coincide [2, 3].

In this letter we investigate the behavior of a non-Hermitian system with parameters which change periodically. We show that in this type of system a new effect *parametric instability near the exceptional point* (*PIEP*) takes place. This effect is the increase of energy in the system due to periodical changing of parameters. This effect resembles parametric resonance in the Hermitian system but there are two fundamental differences. First, in a Hermitian system, energy increases simultaneously with changes in parameters and remains constant otherwise [21]. In the case of *PIEP*, energy increases take place when parameters do not change and all eigenmodes are decaying ($\text{Im} E_1 < 0$, $\text{Im} E_2 < 0$) and energy decreases when parameters change. The second difference comes from the unique properties of the exceptional point. When the system is near the exceptional point, parametric instability occurs when there is almost any change in a parameter, while in the case of a Hermitian system it is necessary to fulfill certain conditions [24].

We illustrate the phenomenon of *PIEP* in the case of two coupling waveguides with gain and loss when the loss in one waveguide is more than the gain in the other. In this system, total energy may increase indefinitely even when all waveguide modes are always decaying. This phenomenon opens a wide range of applications in optics, plasmonics, and optoelectronics, where loss is an inevitable problem and plays a crucial role.

### *Non-orthogonality of eigenstates of non-Hermitian systems: energy oscillations and non-exponential transient behavior*

Before introducing the concept of *PIEP*, we recall the important properties of non-Hermitian systems. For simplicity, we consider a two-dimensional non-Hermitian system with two eigenstates $|\varphi_1\rangle$ and $|\varphi_2\rangle$ (here and later we use Dirac's notation). We use normalization $\langle \varphi_1 | \varphi_1 \rangle = \langle \varphi_2 | \varphi_2 \rangle = 1$.

In a Hermitian system, $\langle \varphi_1 | \varphi_2 \rangle = \langle \varphi_2 | \varphi_1 \rangle = 0$ and $\operatorname{Im} E_1 = \operatorname{Im} E_2 = 0$, so the energy is determined by the energy of each mode and does not depend on the time:

$$E = E_1 |a_1|^2 + E_2 |a_2|^2. \tag{1}$$

In a non-Hermitian system, $\langle \varphi_1 | \varphi_2 \rangle = (\langle \varphi_2 | \varphi_1 \rangle)^* \neq 0$ and the energy of the system is determined not only by the mode amplitude but also by mode overlapping [25]. If the eigenvalues of the system are real, that is, $\operatorname{Im} E_1 = \operatorname{Im} E_2 = 0$, then we have

$$E = E_1 |a_1|^2 + E_2 |a_2|^2 + (E_1 + E_2)|A|\cos(\omega(t - t_\theta) + \theta), \tag{2}$$

where $a_{1,2}$ are initial amplitudes of the eigenstates $|\varphi_{1,2}\rangle$, $\theta = \arg(a_1^* a_2)$, $\omega = |\operatorname{Re}(E_1 - E_2)|$ and $A = a_1^* a_2 \langle \varphi_1 | \varphi_2 \rangle$ (see Supplementary material). We suppose that the origin of time is $t_\theta = \arg\langle \varphi_1 | \varphi_2 \rangle / \omega$. In other words, the energy of the system oscillates with frequency $\omega = |\operatorname{Re}(E_1 - E_2)|$ [2, 22] (see Fig. S1 in Supplementary material).

If all the eigenstates are decaying, that is, $\operatorname{Im} E_1 < 0$ and $\operatorname{Im} E_2 < 0$, then non-Hermitian systems exhibits non-exponential transient behavior [23]. In this case the field amplitude increases compared with the initial value in the first stage and decreases exponentially in the second stage [23] (see Fig. S1 in Supplementary material). In other words, the energy has maxima. We designate the diagonal term in (2) as $E^{av} = E_1 |a_1|^2 + E_2 |a_2|^2$ and the non-diagonal ones in (2) as $E^{osc} \cos(\omega t + \theta)$. Note that $E^{av}$ equals the time-average value of energy.

### *Parametric amplification in the system with non-orthogonal eigenstates*

Energy oscillation, which we describe above, can be used for parametric amplification of the energy. To show this, we expose periodic perturbation in the system, which redistributes the energy between diagonal and non-diagonal terms. Let the perturbation turn on at time $t = t_i$ and turn off at time $t = t_f$ so that the parameters of the system take unperturbed values.

In addition, we require that this perturbation does not change the energy of the system. Then in the case in which $\operatorname{Im} E_1 = \operatorname{Im} E_2 = 0$, from Eq. (2) we have the following relation:

$$E_i^{av} + E_i^{osc} \cos(\omega t_i + \theta_i) = E_f^{av} + E_f^{osc} \cos(\omega t_f + \theta_f), \tag{3}$$

where subscripts $i$ and $f$ correspond to initial and final values. Here $(\omega t_f + \theta_f)$ denotes the phase mismatch between the amplitudes of the first and second eigenmodes after perturbation.

Because oscillations during perturbation have a frequency which differs from $\omega$, the term $\omega t_f$ is not connected with system dynamics and is used for unification notation. The diagonal part of the energy $E_f^{av}$ equals

$$E_f^{av} = E_i^{av} + E_i^{osc} \cos(\omega t_i + \theta_i) - E_f^{osc} \cos(\omega t_f + \theta_f). \tag{4}$$

Now we choose the perturbation such that at initial time $t = t_i$ the non-diagonal part is positive, that is, $\cos(\omega t_i + \theta_i) > 0$, while at $t = t_f$ it is negative, that is, $\cos(\omega t_f + \theta_f) < 0$. In such case the diagonal part of the energy at the end of the perturbation will be larger than that at the begin of the perturbation. Note that at the same time the total energy of the system does not change (see Fig. 1). The most favorable case for increasing the diagonal part of the energy is when $\cos(\omega t_i + \theta_i) = 1$ and $\cos(\omega t_f + \theta_f) = -1$. In this case the diagonal part of the energy after perturbation $E_f^{av} = E_i^{av} + E_i^{osc} + E_f^{osc}$ is larger than both the diagonal part of the energy $E_i^{av}$ and the highest possible value of the total energy of the system $E_i^{av} + E_i^{osc}$ before perturbation. In other words, after perturbation, the diagonal part of the energy and the highest possible value of the total energy of the system increase. Now if we repeat such a perturbation with the period which is equal to an odd number of half periods of energy oscillation, e.g. $\Delta = T/2 = \pi / |\mathrm{Re}(E_1 - E_2)|$, we will have permanent growth of the energy in the system (Fig. 1).

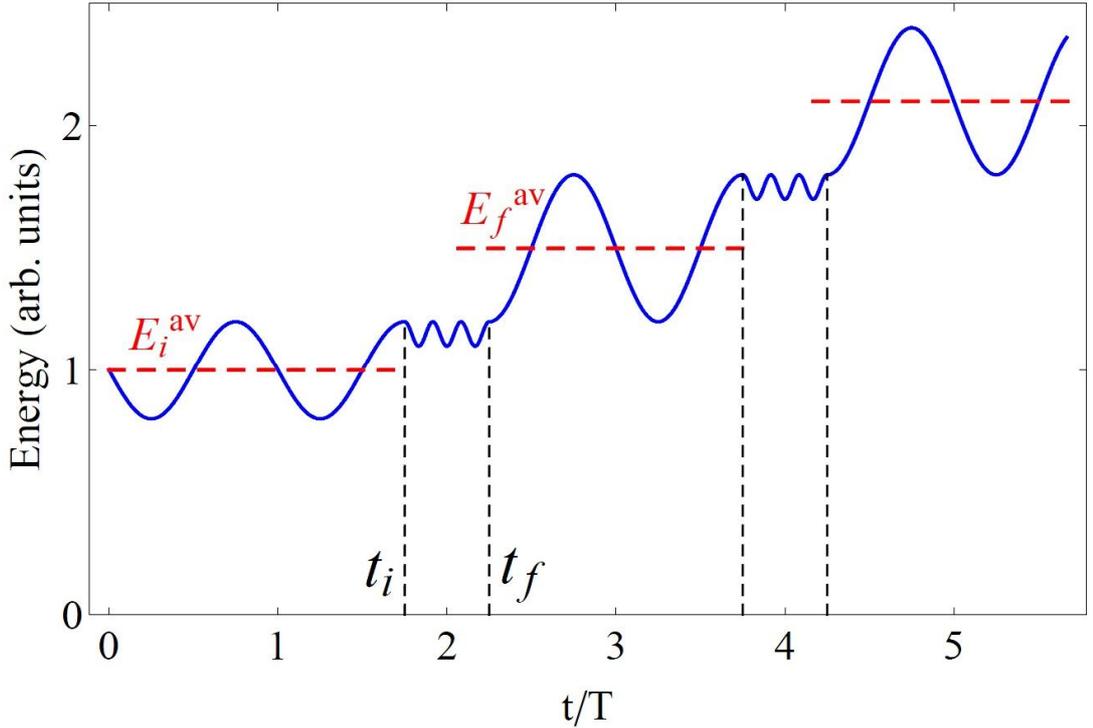

FIG. 1. (Color online) The dependence of the energy in a system with perturbation on time. Black dashed vertical lines denote the start and end times of the perturbations. Red horizontal dashed lines denote the average value of the energy (diagonal part) between perturbations.

The growth of the energy in the system may take place even when all the eigenvalues are constant and have negative imaginary parts, that is, $\mathrm{Im}\, E_1 < 0$, $\mathrm{Im}\, E_2 < 0$ (see Fig. S2 in Supplementary material). In this case perturbation in the system redistributes the energy between diagonal and non-diagonal terms, as in the previous case. The total energy of the system increases in the time interval when all the eigenvalues are constant and have negative imaginary parts ($\mathrm{Im}\, E_1 < 0$, $\mathrm{Im}\, E_2 < 0$), while during perturbation, energy decreases. This statement points out the first difference between parametric instability in a non-Hermitian system and parametric resonance in a Hermitian one where energy increases simultaneously with changes in parameters.

*Describing suitable perturbation*

As we mentioned above, to realize the increase in energy in the system it is necessary that $\cos(\omega t_i + \theta_i) \approx 1$ and $\cos(\omega t_f + \theta_f) \approx -1$. In a non-Hermitian system such perturbation is simply realized near EP, where two eigenstates almost coincide with each other, $|\varphi_1\rangle \approx |\varphi_2\rangle$. Let us introduce the normalized state $|\varphi_\perp\rangle$ which is orthogonal to that of the eigenstate: $\langle \varphi_1 | \varphi_\perp \rangle = 0$, $\langle \varphi_\perp | \varphi_\perp \rangle = 1$. Then the second eigenstate $|\varphi_2\rangle$ can be written in the form $|\varphi_2\rangle = \sqrt{1-c^*c}|\varphi_1\rangle + c|\varphi_\perp\rangle$, where $c = \langle \varphi_\perp | \varphi_2 \rangle$. Note that near the EP we have $|c| \ll 1$ (see Supplementary material). In time $t = t_i$, let the system state be

$$|\psi_i\rangle = a_1|\varphi_1\rangle + a_2|\varphi_2\rangle = a_1\left(|\varphi_1\rangle + \left|\frac{a_2}{a_1}\right|\exp(i\theta)|\varphi_2\rangle\right), \tag{5}$$

where $\theta = \arg(a_2/a_1) = \arg(a_2 a_1^*/|a_1|^2) = \arg(a_2 a_1^*)$. Note that we have the same phase $\theta$ as in Eq. (2).

Consider an external perturbation that changes the system parameters such that in time $t = t_f$ the system state takes the form

$$|\psi_f\rangle = b_1|\varphi_1\rangle + b_2|\varphi_\perp\rangle. \tag{6}$$

Expansion $|\psi_f\rangle$ in the system eigenstate gives

$$|\psi_f\rangle = \left(b_1 - \frac{b_2}{c}\sqrt{1-c^*c}\right)|\varphi_1\rangle + \frac{b_2}{c}|\varphi_2\rangle. \tag{7}$$

From (7) we see that when perturbation is such that $|b_2/b_1| \gg |c|$, the amplitudes of the eigenstates after perturbation (expansion coefficients of $|\psi_f\rangle$ in (7)) are approximately equal by modulus and have opposite signs:

$$\left(b_1 - \frac{b_2}{c}\sqrt{1-c^*c}\right) \simeq -\frac{b_2}{c}. \tag{8}$$

Equations (8) show that the final state corresponds to the phases $\cos(\omega t_f + \theta_f) \approx -1$. So, if the chosen the initial state corresponds to the phases $\cos(\omega t_i + \theta_i) \approx 1$, then perturbation satisfies all necessary conditions for observation of energy amplification. Note that after a half period of energy oscillation, $\Delta = T/2 = \pi/|\text{Re}(E_1 - E_2)|$, we have $\cos(\omega t + \theta) = 1$ and perturbation may be repeated. Here we point out the second difference between parametric resonance and parametric instability. Note that near *EP*, $c = \langle \varphi_\perp | \varphi_2 \rangle \approx 0$ and condition $|b_2/b_1| \gg |c|$ are satisfied automatically! This means that almost any perturbation results in increasing energy, while in the case of parametric resonance in the Hermitian system it is necessary to fulfill certain conditions [24]. This key point allows us to consider this effect as a new phenomenon: *parametric instability near the exceptional point* (*PIEP*).

It is very important to emphasize that the non-Hermitian character of the system is crucial. Indeed, in the Hermitian system all eigenstates are orthogonal ($c = \langle \varphi_\perp | \varphi_2 \rangle = 1$) and the energy of the system does not depend on the phase mismatch between the amplitudes of eigenstates because of the orthogonality of the eigenstates.

To sum up, in a non-Hermitian system, a certain change of the system parameters leads to increasing energy. Note that in EP it happens at any change. Such an increase arises due to energy transfer from the oscillating non-diagonal part to the diagonal part. As a result, the diagonal part of the energy and the amplitude of the eigenstates also increase.

If we change the system parameters periodically with the period which is equal to an odd number of half periods of energy oscillation, e.g. $\Delta = T/2 = \pi/|\text{Re}(E_1 - E_2)|$, then we will have permanent growth of the energy, which is limited only by nonlinear effects.

### PIEP in an optical system

Above we define the conditions at which *PIEP* takes place. Now we give an example of a system with *PIEP*. Let us consider a system consisting of two coupling waveguides in which the real parts of the waveguides refractive index are equal, that is, $\text{Re}\,n_1 = \text{Re}\,n_2$, while the imaginary parts are different and have opposite signs, that is, $\text{Im}\,n_1 > 0$, $\text{Im}\,n_2 < 0$, and loss exceeds gain, $|\text{Im}\,n_1| \geq |\text{Im}\,n_2|$ [14], and the distance between waveguides changes with the period see Fig. 2.

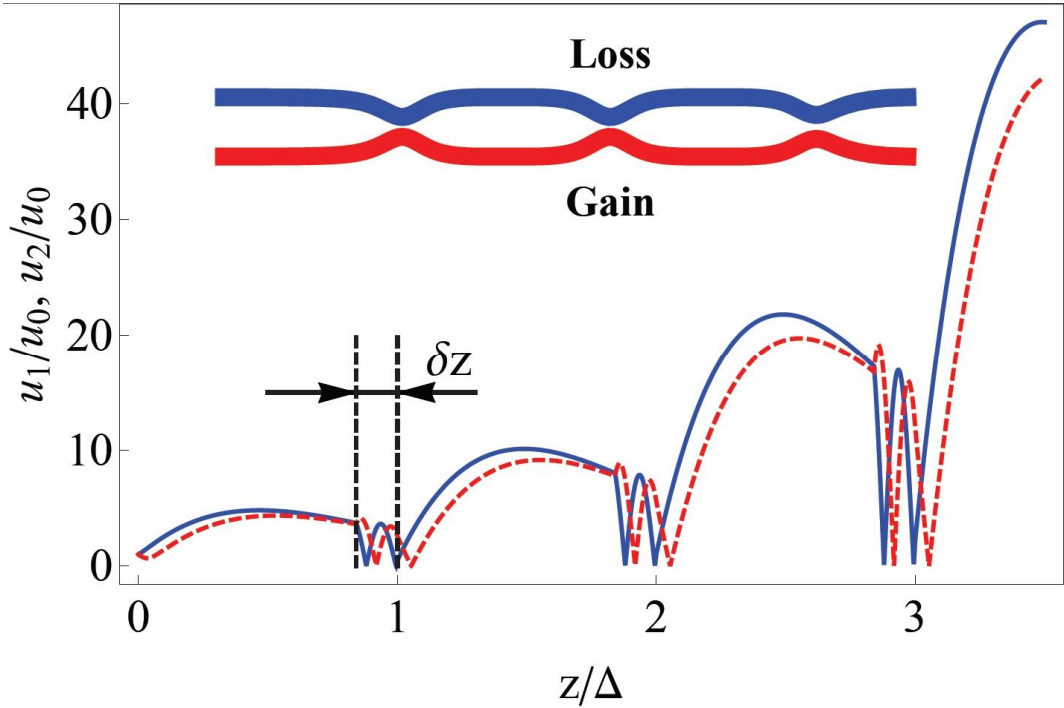

FIG. 2. (Color online) The dependence of field intensity $u_1/u_0$ in the first waveguide (blue solid line) and $u_2/u_0$ in the second one (red dashed line) on the coordinate along waveguides $z$. $u_0 = u_1(z=0) = u_2(z=0)$ is the initial amplitude in both waveguides. The distance between waveguides changes with the period $\Delta = T/2 = \pi/|\text{Re}(E_1 - E_2)|$, perturbation length $\delta z \ll \Delta$, $g^2 = 2.5 \cdot 10^{-3}$, $\gamma = 5 \cdot 10^{-3}$ and $|\kappa|^2 = 1.01 g^2$ outside the perturbation region and $|\kappa|^2 = 4g^2$ inside the perturbation region. Inset: system under consideration.

Let the z-axis be directed along the waveguides and let the amplitudes of the waveguides electric fields be $u_1$ and $u_2$. In coupled-mode theory [1, 14, 25], the dependence of the field amplitude on the coordinate z while it propagates is described by the following system:

$$-i\frac{d}{dz}\begin{pmatrix}u_1\\u_2\end{pmatrix} = \begin{pmatrix}\beta + i\gamma + ig & \kappa^*(z)\\ \kappa(z) & \beta + i\gamma - ig\end{pmatrix}\begin{pmatrix}u_1\\u_2\end{pmatrix}, \qquad (9)$$

where $\beta$ is the real part of the wavenumbers. For symmetry, we introduce half of the sum of the imaginary parts of the wavenumbers $\gamma$ (which corresponds to common background damping or amplification) and half of the difference between the imaginary parts $g$. $\kappa(z)$ is the coupling constant, which depends on the distance between waveguides. Equations (9) for the system of two coupled waveguides are equivalent to the Schrodinger equation for the two-level non-Hermitian system with the replacement of time $t$ at inversion coordinate $z$.

When the coupling constant $\kappa$ does not depend on coordinate $z$, the system energy $E = |u_1|^2 + |u_2|^2$ non-exponentially depends on the distance $z$: at first the energy grows due to non-orthogonal eigenmodes of the system; then this growth is changed by the exponential decay, and as a result the energy tends to zero when $z \to \infty$; see also Fig. S1 in Supplementary material and [23].

We consider the case when the distance between waveguides is constant except for the region with length $\delta z$ (a perturbation length), which are repeated with period $\Delta \gg \delta z$. In this situation parametric instability near the exceptional point may be observed. PIEP leads to the growth of energy in the system (Fig. 2). The energy growth may occur even if we choose the values of system parameters at which eigenmodes are always decaying. For example, imaginary parts of the eigenvalues of the system from Fig. 2 is always positive and the eigenmodes are decaying (see Supplementary material). Note that energy before perturbation $E_i$ is no less than energy after perturbation $E_f$, see Fig. 2. That points out the difference between PIEP and ordinary parametric resonance in the Hermitian system.

To observe PIEP, it is necessary that before perturbation the state has the value $\cos(\omega z_i + \theta_i) \approx 1$, while after perturbation the state has the value $\cos(\omega z_f + \theta_f) \approx -1$. Here $\omega z + \theta$ is the phase difference between the amplitudes of eigenmodes.

In the previous section, it has been shown that near the EP any perturbation is suitable. This statement is confirmed by numerical simulation: condition $\cos(\omega z_f + \theta_f) \approx -1$ is satisfied for a wide range of perturbation lengths $\delta z$ (Fig. 3). Moreover, when the system parameters tend to EP, the suitable perturbation length increases (Fig. 3).

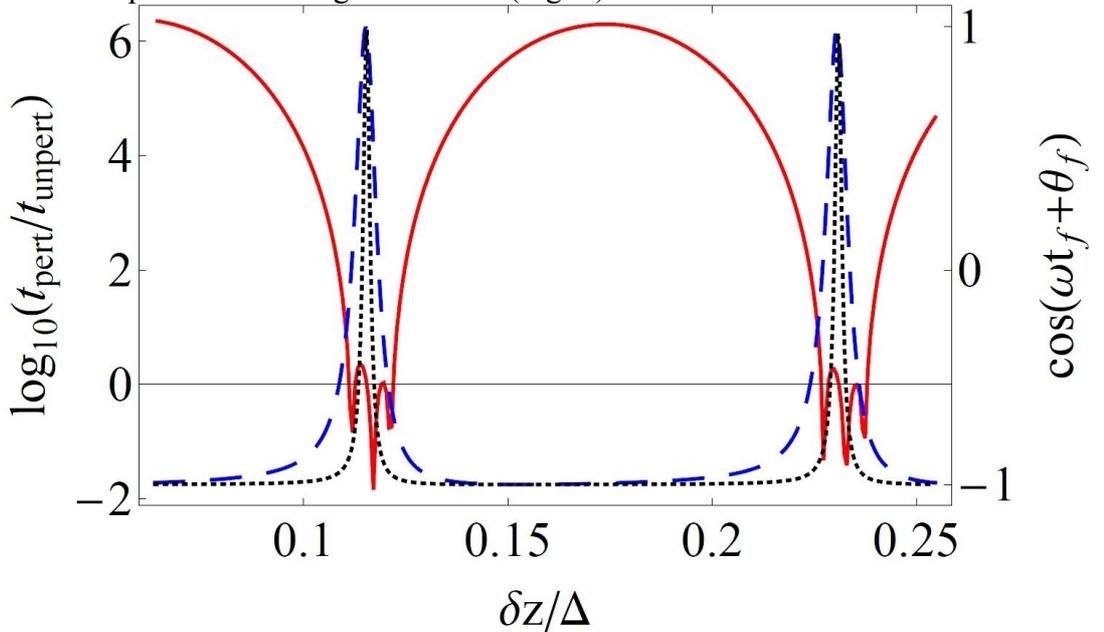

FIG. 3. (Color online) Red solid line: the dependence of the ratio (logarithmic scale) of the transmission coefficient of perturbed $t_{pert}$ to unperturbed $t_{unpert}$ systems on the perturbation length $\delta z$. Blue dashed line: the dependence of the value $\cos(\omega z_f + \theta_f)$ after perturbation on the perturbation length $\delta z$. For both lines (red and blue) $|\kappa|^2 = 1.01 g^2$. Black dotted line: the dependence of the value $\cos(\omega z_f + \theta_f)$ after perturbation on the perturbation length $\delta z$, in the case of $|\kappa|^2 = 1.001 g^2$. EP corresponds to the case when $|\kappa|^2 = g^2$. Other parameters are the same as in Fig. 2. The initial state is $|\psi_i\rangle = a(|\varphi_1\rangle + |\varphi_2\rangle)$.

For a perturbation length $\delta z$ for which $\cos(\omega z_f + \theta_f) \approx -1$, the transmission coefficient of the perturbed system is larger than that of the unperturbed system (Fig. 3).

The opposite situation is realized only for a small region $\delta z$ at which $\cos(\omega z_f + \theta_f) \approx 1$ (see Fig. 3). It is important that the transmission coefficient of the perturbed system is larger even if the period of perturbation $\Delta_{var}$ does not coincide with its optimal value $\Delta = T/2 = \pi / |\mathrm{Re}(E_1 - E_2)|$ (Fig. 4). From Fig. 4 it is seen that energy amplification takes place at almost any perturbation. So, it is possible for the energy to be unlimitedly amplified in the system where loss is dominated by gain and all eigenmodes are decreasing. This effect is connected with the new physical phenomenon, *parametric instability near the exceptional point*, which was described in the previous sections.

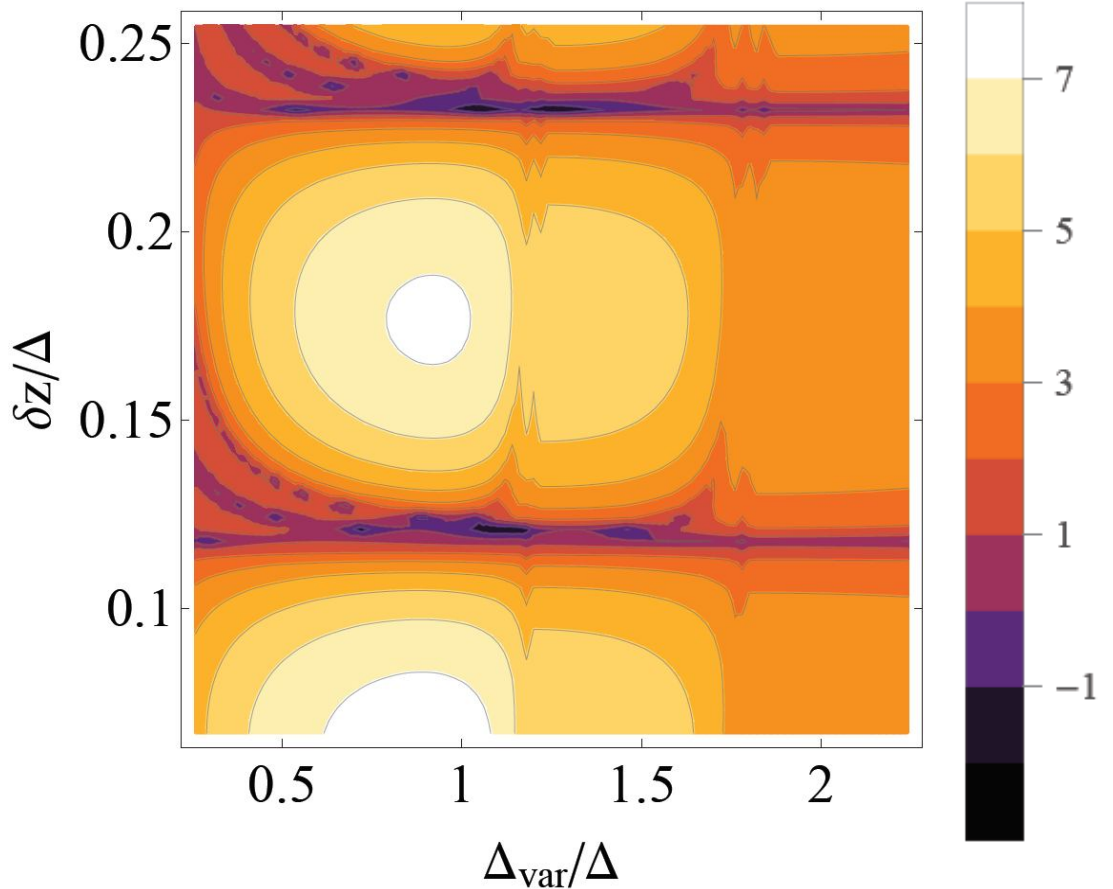

FIG. 4. (Color online) The dependence of the ratio (logarithmic scale) of the transmission coefficient of perturbed $t_{pert}$ to unperturbed $t_{unpert}$ systems on the perturbation period $\Delta_{var}$ (in units of the optimal period $\Delta = T/2 = \pi / |\mathrm{Re}(E_1 - E_2)|$ and the perturbation length $\delta z$, $|\kappa|^2 = 1.001 g^2$. Other parameters are the same as in Fig. 2.

All previous consideration concerns a linear system. However, when the amplitude $u_{1,2}$ increases it is necessary to take into account nonlinear effects which are connected with saturation of the active medium. *PIEP* takes place in this case too (see Supplementary). The difference is that the energy grows to some finite value (see Fig. S3 in Supplementary material), which depends on the perturbation parameters $\delta z$ and $\kappa$. More careful analysis we give in forthcoming paper.

*Conclusion*

In non-Hermitian systems, eigenstates are non-orthogonal. The maximal degree of non-orthogonality corresponds to EP, where several eigenstates coincide. In the present work, we show that non-orthogonality allows us to observe a new effect, *parametric instability near the exceptional point* (*PIEP*), which results in increasing energy when the system parameters are periodically changed. This increase takes place even when all system eigenstates are decaying. This effect has two features which distinguish it from parametric resonance in a Hermitian system. First, in a Hermitian system, energy increases simultaneously with changes in parameters and remains constant otherwise. In the case of *PIEP*, on one hand, an increase in energy takes place when parameters do not change and all eigenstates are decaying, and on the other hand, a decrease in energy takes place when the parameters change. The second difference comes from the unique properties of the exceptional point. When system is near EP, parametric instability occurs with almost any parameter change, while in the case of Hermitian systems it is necessary to fulfill certain conditions [24].

As an example we consider two waveguides with loss and gain and a coupling constant which is periodically perturbed. This system manifests *PIEP*. Changing the coupling constant leads to increases in energy which are limited only by nonlinear effects. Moreover, we show that the transmission coefficient of such waveguides is larger that of a system with constant parameters. Near EP we have a wide range of changes in parameters at which parametric instability occurs (Fig. 4). These results correspond to the feature of parametric instability which we mentioned above.

The phenomenon of *PIEP* may be used in metamaterial, plasmonic, and nanooptic devices where applicability is sufficiently restricted by losses.

The authors thank Yu. E. Lozovik and N. M. Shchelkachev for a helpful discussion. This work was supported in part by the Advanced Research Foundation (Contract No. 7/004/2013-2018), RFBR grant nos. 12-02-010193 and 13-02-00407, as well as by the "Dynasty" foundation.

[1] R. El-Ganainy et al., *Opt. Lett.*, **32**, 2632 (2007).
[2] K. Makris et al., *Phys. Rev. Lett.*, **100**, 103904 (2008).
[3] S. Klaiman et al., *Phys. Rev. Lett.*, **101**, 080402 (2008).
[4] C. M. Bender and S. Boettcher, *Phys. Rev. Lett.*, **80**, 5243 (1998).
[5] C. M. Bender , et al., *Phys. Rev. Lett.*, **89**, 270401 (2002).
[6] C. M. Bender, *Reports on Progress in Physics*, **70**, 947 (2007).
[7] S. Longhi, *Phys. Rev. Lett.*, **103**, 123601 (2009).
[8] Y. Chong et al., *Phys. Rev. Lett.*, **106**, 093902 (2011).
[9] Z. Lin et al., *Phys. Rev. Lett.*, **106**, 213901 (2011).
[10] M. Liertzer et al., *Phys. Rev. Lett.*, **108**, 173901 (2012).
[11] N. Chtchelkatchev et al., *Phys. Rev. Lett.*, **109**, 150405 (2012).
[12] A. A. Zyablovsky et al., *Phys. Rev. A*, **89**, 033808 (2014).
[13] A. A. Zyablovsky et al., *Physics-Uspekhi*, **57**, 1063 (2014).
[14] A. Guo et al., *Phys. Rev. Lett.*, **103**, 093902 (2009).
[15] C. E. Ruter et al., *Nature Physics*, **6**, 192 (2010).
[16] L. Feng et al., *Science*, **333**, 729 (2011).

[17] A. Regensburger et al., *Nature*, **488**, 167 (2012).
[18] L. Feng et al., *Nature Materials*, **12**, 108 (2013).
[19] B. Peng et al., *Nature Physics*, **10**, 394 (2014).
[20] H. Hodaei et al., *Science*, **346**, 975 (2014).
[21] R. Fleury et al, *Nature Commun.*, **6**, 5905 (2015).
[22] B. Bagchi et al., Mod. Phys. Lett. A, **16**, 2047 (2001).
[23] K. Makris et al., *Phys. Rev. X*, **4**, 041044 (2014).
[24] L. D. Landau and E. M. Lifshitz, *Course Of Theoretical Physics: Volume 1 Mechanics*, (Butterworth-Heinemann, 1976).
[25] K. Okamoto, *Fundamentals of optical waveguides* (Academic press, 2010).

# Supplementary material for "Parametric instability of non-Hermitian systems near the exceptional point"

A.A. Zyablovsky,[1,2] E.S. Andrianov,[1,2] A.A. Pukhov[1,2,3]

[1] All-Russia Research Institute of Automatics, 22 Suschevskaya, Moscow, Russia
[2] Moscow Institute of Physics and Technology, 9 Institutsky per., Dolgoprudny, Russia
[3] Institute Theoretical and Applied Electromagnetics, 13 Izhorskay per., Moscow, Russia

*Non-orthogonality of eigenstates of non-Hermitian systems: energy oscillations and non-exponential transient behavior*

Non-orthogonality of eigenstates in a non-Hermitian system leads to energy oscillation [2, 22]. This phenomenon is the consequence of the dependence of the system energy on both amplitude and (in contrast to a Hermitian system) the phase of eigenstates. For simplicity, we consider a two-dimensional non-Hermitian system with two eigenstates $|\varphi_1\rangle$ and $|\varphi_2\rangle$ (here and later we use Dirac's notation). Then the system state after time $t$ has the form

$$|\psi\rangle = a_1 \cdot \exp(-iE_1 t)|\varphi_1\rangle + a_2 \cdot \exp(-iE_2 t)|\varphi_2\rangle, \quad (S1)$$

where $a_i$ and $E_i$ are initial amplitudes and eigenvalues of the eigenstates $|\varphi_i\rangle$, respectively. To determine the energy of the system we introduce the system Hamiltonian $\hat{H} = E_1|\varphi_1\rangle\langle\varphi_1| + E_2|\varphi_2\rangle\langle\varphi_2|$. Then the energy of the system is given by the equation

$$\begin{aligned} E = \langle\psi|\hat{H}|\psi\rangle = & E_1|a_1|^2 \exp(2\operatorname{Im}(E_1)t) + E_2|a_2|^2 \exp(2\operatorname{Im}(E_2)t) + \\ & + E_2 a_1^* a_2 \cdot \exp(i\operatorname{Re}(E_1 - E_2)t)\exp(\operatorname{Im}(E_1 + E_2)t)\langle\varphi_1|\varphi_2\rangle + \\ & + E_1 a_1 a_2^* \cdot \exp(-i\operatorname{Re}(E_1 - E_2)t)\exp(\operatorname{Im}(E_1 + E_2)t)\langle\varphi_2|\varphi_1\rangle \end{aligned} \quad (S2)$$

Here we use normalization $\langle\varphi_1|\varphi_1\rangle = \langle\varphi_2|\varphi_2\rangle = 1$.

In a Hermitian system, $\langle\varphi_1|\varphi_2\rangle = \langle\varphi_2|\varphi_1\rangle = 0$ and $\operatorname{Im} E_1 = \operatorname{Im} E_2 = 0$, so the energy is determined by the energy of each mode and does not depend on the time:

$$E = E_1|a_1|^2 + E_2|a_2|^2. \quad (S3)$$

In a non-Hermitian system, $\langle\varphi_1|\varphi_2\rangle = (\langle\varphi_2|\varphi_1\rangle)^* \neq 0$ and the energy of the system is determined not only by the mode amplitude but also by mode overlapping. This energy oscillates with frequency $\omega = |\operatorname{Re}(E_1 - E_2)|$ [2, 22].

We show later that these time oscillations can be used for parametric amplification of the field in the system without an external source. Such amplification is realized due to the non-Hermitian character of the system.

For simplicity, we rewrite Eq. (S2) in the form

$$\begin{aligned} E = & E_1|a_1|^2 \exp(2\operatorname{Im}(E_1)t) + E_2|a_2|^2 \exp(2\operatorname{Im}(E_2)t) + \\ & + (E_2 A \exp(i\omega t) + E_1 A^* \exp(-i\omega t))\exp(\operatorname{Im}(E_1 + E_2)t) \end{aligned}, \quad (S4)$$

where $A = a_1^* a_2 \langle\varphi_1|\varphi_2\rangle$. The real part of $E$ describes the energy of the system while the imaginary part of $E$ corresponds to dissipation of the system.

Behavior of the system strongly depends on the value of the system parameters. If the eigenvalues of the system are real, that is, $\operatorname{Im} E_1 = \operatorname{Im} E_2 = 0$ (as in a PT-symmetrical system below the exceptional point), then we have

$$\operatorname{Re} E = E_1|a_1|^2 + E_2|a_2|^2 + (E_1 + E_2)|A|\cos(\omega(t - t_\theta) + \theta), \quad (S5)$$

where $\theta = \arg(a_1^* a_2)$ and $\omega t_\theta = \arg\langle \varphi_1 | \varphi_2 \rangle$. In other words, the energy of the system oscillates with time (Fig. S1) [2].

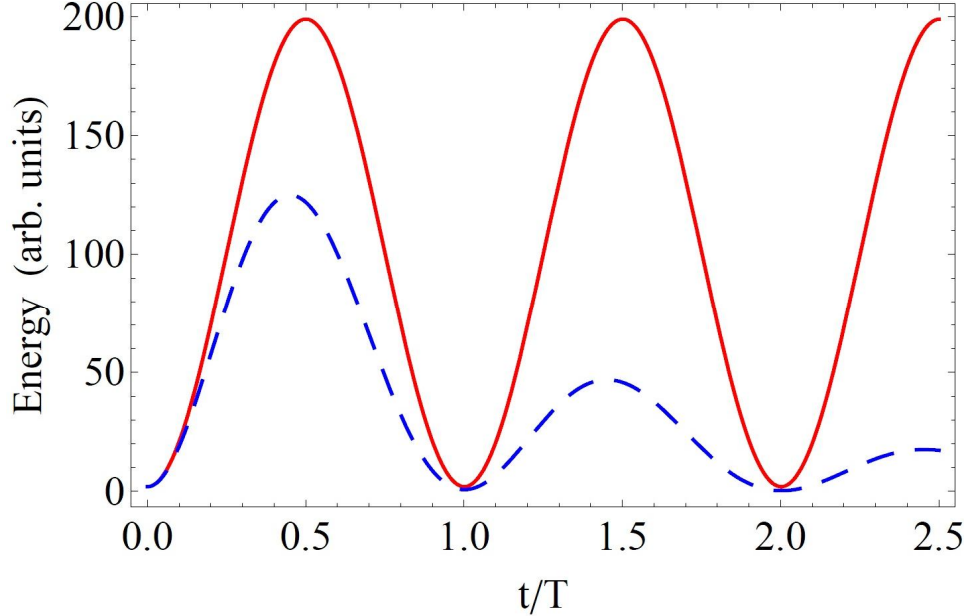

FIG. S1. (Color online) (a) Red solid line: energy oscillation in the non-Hermitian system when its eigenvalues are real; blue dashed line: non-exponential transient behavior in the non-Hermitian system when its eigenvalues are complex. $T = 2\pi / |\mathrm{Re}(E_1 - E_2)|$ is the oscillation period.

If all the eigenstates are decaying, that is, $\mathrm{Im}\, E_1 < 0$ and $\mathrm{Im}\, E_2 < 0$, then

$$\mathrm{Re}\, E = \mathrm{Re}\, E_1 |a_1|^2 \exp(2\,\mathrm{Im}(E_1)t) + \mathrm{Re}\, E_2 |a_2|^2 \exp(2\,\mathrm{Im}(E_2)t) + \qquad (S6)$$
$$+ \mathrm{Re}(E_1 + E_2) |A| \cos(\omega(t - t_\theta) + \theta) \exp(\mathrm{Im}(E_1 + E_2)t),$$

where $\theta = \arg(a_1^* a_2)$ and $\omega t_\theta = \arg\langle \varphi_1 | \varphi_2 \rangle$. We see that the system exhibits non-exponential transient behavior [23]. In other words, the energy has maxima (Fig. S1). Note that the growth of the energy in the system may take place even in this case, see Fig. S2.

*Properties of exceptional point*

At the exceptional point, the spectrum of the system is degenerate, that is, $E_1 = E_2 = E$, and the eigenstates are equal to each other, that is, $|\varphi_1\rangle = |\varphi_2\rangle = |\varphi\rangle$; i.e. they do not form the basis [26]. In other words, at this point they have the maximal degree of non-orthogonality. To construct the basis it is necessary to add to the eigenstate the adjoined one, $|\varphi_{adj}\rangle$, which is determined by the equation [4]

$$(\hat{H} - E \cdot \hat{I})|\varphi_{adj}\rangle = |\varphi\rangle, \qquad (S7)$$

and the time evolution of states of the system has the form

$$|\psi(t)\rangle = (a_{adj} t + a_1) e^{-iEt} |\varphi\rangle + a_{adj} e^{-iEt} |\varphi_{adj}\rangle. \qquad (S8)$$

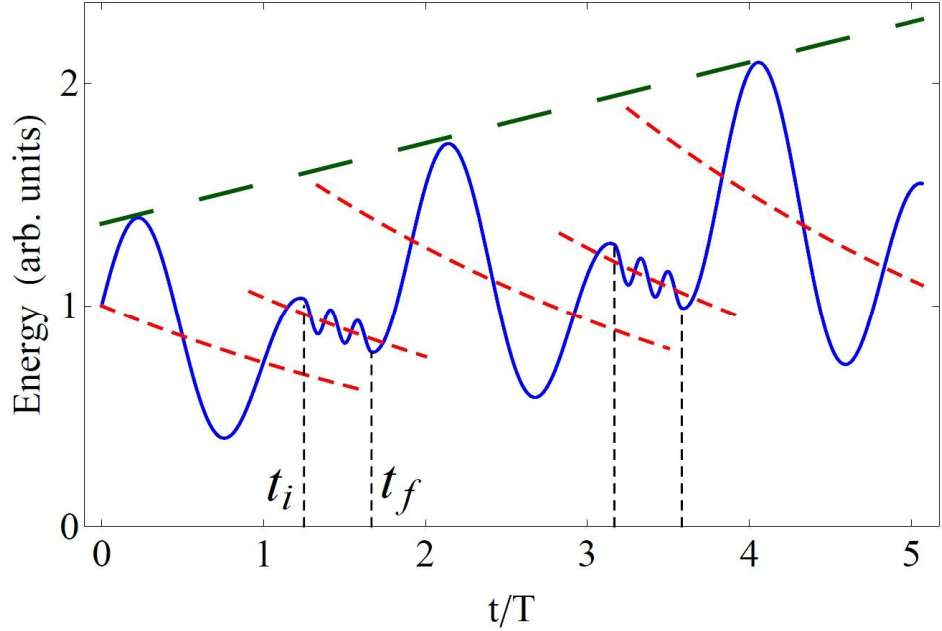

FIG. S2. (Color online) The dependence of the energy in a system with perturbation on time. Black dashed vertical lines denote the start and end of the perturbations. Red horizontal dashed lines denote the diagonal part of the energy between perturbations. $\text{Im} E_1 < 0$, $\text{Im} E_2 < 0$

*Eigenvalues and eigenmodes*

The dependence of the field on the coordinate $z$ in the system with $\kappa$ does not depend on coordinate $z$, as follows:

$$\begin{pmatrix} u_1 \\ u_2 \end{pmatrix} = a_1 \exp(iE_1 z)|\varphi_1\rangle + a_2 \exp(iE_2 z)|\varphi_2\rangle. \tag{S9}$$

where constants $u_1$ and $u_2$ are determined by the initial conditions. The eigenvalues of the system (S9) are

$$E_{1,2} = \beta + i\gamma \pm \sqrt{|\kappa|^2 - g^2}. \tag{S10}$$

and their eigenmodes (waveguide modes) are

$$|\varphi_{1,2}\rangle = \begin{pmatrix} 1 \\ \dfrac{-ig \pm \sqrt{|\kappa|^2 - g^2}}{\kappa^*} \end{pmatrix}. \tag{S11}$$

Note that in this case, decaying eigenmodes have positive imaginary parts, that is, $\text{Im} E_{1,2} > 0$.

If $|\kappa|^2 > g^2$ then the imaginary parts of all eigenvalues are positive, $\text{Im} E_{1,2} > 0$ and all eigenmodes are decay. In the main text we consider this case because the condition $|\kappa|^2 > g^2$ is satisfied at all $z$, Fig. S3. So all eigenmodes of coupled waveguides considered in the text are decay.

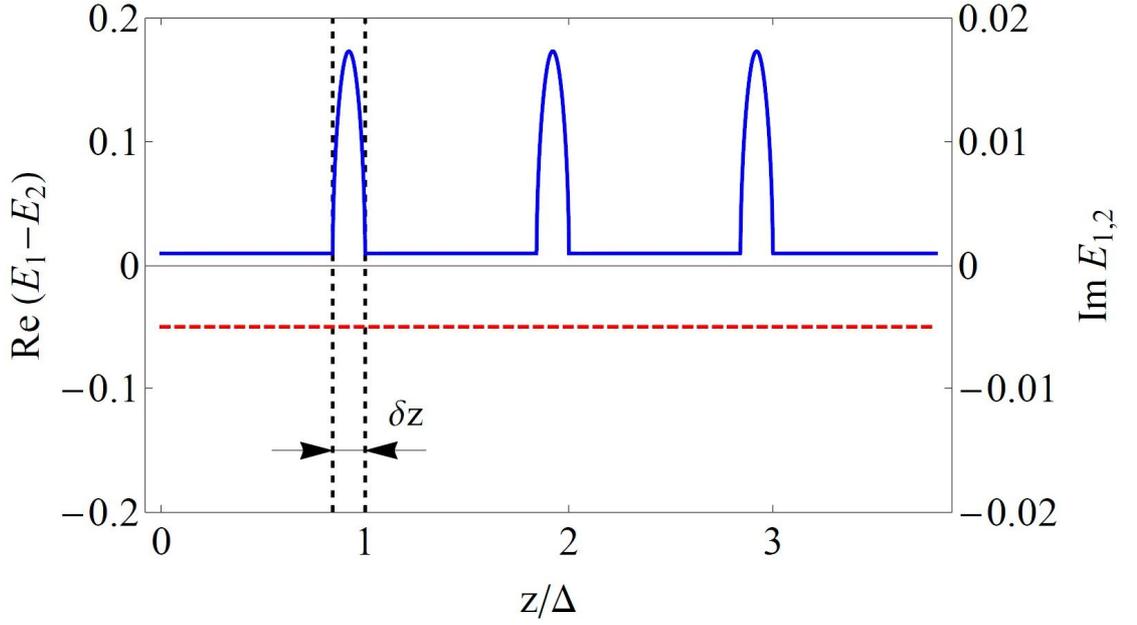

Fig. S3. (Color online) The difference between the real parts of eigenvalues (blue solid line); Imaginary parts (red dashed line) of eigenvalues. For both parts of the figure, $g^2 = 2.5 \cdot 10^{-3}$, $\gamma = 5 \cdot 10^{-3}$, and $\kappa(z)$ changes as in Fig. 2.

*Nonlinear effects*

When the amplitude of the waveguides electric fields $u_{1,2}$ increases it is necessary to take into account nonlinear effects which are connected with saturation of the active medium. To do this, let us suppose that amplification and losses in each waveguide depend on the field amplitude in the following [27, 28]

$$g_{1,2} = (-1)^{1,2} \frac{g_c}{1+\alpha |u_{1,2}|^2}. \tag{S12}$$

where $\alpha$ is the nonlinearity coefficient. As follows from the results of numerical simulation *PIEP* takes place in this case too (Fig. S4).

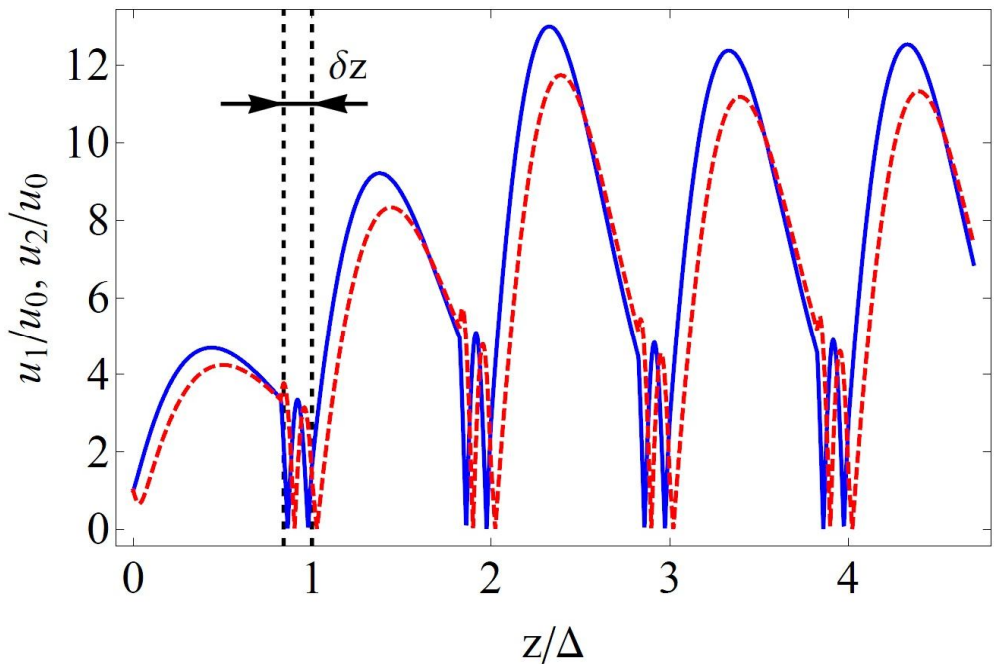

Fig. S4. (Color online) The dependence of field amplitudes $u_1/u_0$ and $u_2/u_0$ in the nonlinear

waveguides (see Eq. S12) on the coordinate along waveguides $z$ (in units of the optimal period $\Delta = \pi / |\mathrm{Re}(E_1 - E_2)|$). Here $u_0 = u_1(z=0) = u_2(z=0)$ is the initial amplitude. The parameters are as follows: $g_{1,2} = (-1)^{1,2} \dfrac{g_c}{1+\alpha |u_{1,2}|^2}$, $g_c = 0.05$, $\gamma = 5 \cdot 10^{-3}$, $\Delta = \pi / |\mathrm{Re}(E_1 - E_2)|$, and $\alpha = 10^{-4}$.

[26] T. Kato, *Perturbation theory of linear operators* (Springer, 1966).
[27] A. E. Miroshnichenko et al., *Phys. Rev. A*, **84**, 012123 (2011).
[28] A. V. Dorofeenko et al., *Physics-Uspekhi*, **55**, 1080 (2012).